\documentclass[11pt]{article}

\usepackage[final]{acl}
\usepackage{times}
\usepackage{latexsym}
\usepackage{amsmath}

\usepackage[T1]{fontenc}

\usepackage[utf8]{inputenc}

\usepackage{microtype}

\usepackage{inconsolata}

\usepackage{graphicx}
\usepackage[normalem]{ulem}
\usepackage{amssymb}
\title{BioHiCL: Hierarchical Multi-Label Contrastive Learning for Biomedical Retrieval with MeSH Labels}

\author{Mengfei Lan, Lecheng Zheng, Halil Kilicoglu\thanks{Corresponding author} \\
  University of Illinois Urbana-Champaign \\
  \texttt{\{mlan3, lecheng4, halil\}@illinois.edu} \\}

\begin{document}
\maketitle
\begin{abstract}
Effective biomedical information retrieval requires modeling domain semantics and hierarchical relationships among biomedical texts. Existing biomedical generative retrievers build on coarse binary relevance signals, limiting their ability to capture semantic overlap. We propose BioHiCL (\textit{Biomedical Retrieval with Hierarchical Multi-Label Contrastive Learning}), which leverages hierarchical MeSH annotations to provide structured supervision for multi-label contrastive learning. Our models, BioHiCL-Base (0.1B)\footnote{https://huggingface.co/LunaLan07/BioHiCL-Base} and BioHiCL-Large (0.3B)\footnote{https://huggingface.co/LunaLan07/BioHiCL-Large}, achieve promising performance on biomedical retrieval, sentence similarity, and question answering tasks, while remaining computationally efficient for deployment.

\end{abstract}
\vspace{-2mm}
\section{Introduction}
\vspace{-2mm}
Dense retrievers learn vector representations of text from large-scale unsupervised, supervised, and synthetic corpora, achieving strong performance on general-domain information retrieval (IR) benchmarks~\cite{ni2022large, wang-etal-2024-improving-text, xiao2024c, lassance2024splade}. However, these models often fail to capture biomedical-specific semantics, limiting their effectiveness in tasks involving specialized terminology.

To bridge this gap, various biomedical IR models have been proposed. Encoder-based biomedical language models, such as BiomedBERT~\cite{chakraborty2020biomedbert}, pretrain transformer encoders on large-scale biomedical corpora to capture domain-specific terminology and semantics, but their retrieval effectiveness is limited by the availability of retrieval task-specific supervision. More recently, large-scale biomedical IR models, including MedCPT and BMRetriever~\cite{jin2023medcpt, xu2024bmretriever}, leverage contrastive learning on extensive biomedical semantic-related text pairs to achieve improved semantic alignment. Despite their effectiveness, these models depend on computationally expensive training and coarse-grained relevance signals (e.g., MedNLI-derived entailment labels~\cite{xu2024bmretriever} or query–article clicks~\cite{jin2023medcpt}), making it challenging to capture the fine-grained and partially overlapping semantics characteristic of biomedical texts.

\begin{figure}[t]
\includegraphics[width=7cm]{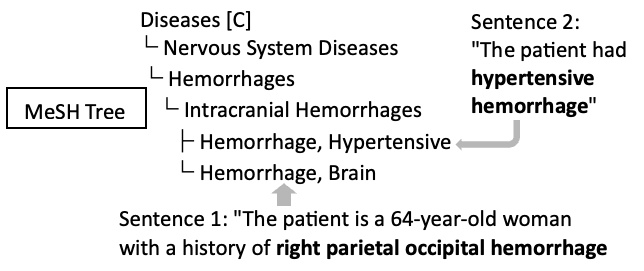}
\centering
\vspace{-4mm}
\caption{The MeSH labels for a pair of sentences annotated as "neutral" (irrelevant) in MedNLI.}
\label{introduction_sample}
\vspace{-6.5mm}
\end{figure}

Semantic relationships in biomedical text are complex. Documents and sentences often share partial semantic overlap that cannot be adequately modeled with binary relevance signals alone, limiting the ability of dense retrievers to learn fine-grained biomedical representations. Medical Subject Headings (MeSH) offer a natural source of multi-aspect supervision. MeSH is a hierarchically organized controlled vocabulary curated by domain experts~\cite{nlm_mesh}, explicitly encoding both concept overlap and hierarchical structure. 
For example, Figure~\ref{introduction_sample} shows a sentence pair labeled as \emph{neutral} in MedNLI, while their MeSH annotations share a common parent concept (\textit{Intracranial Hemorrhages}) in the disease hierarchy indicating the relevance between sentences. This hierarchy-aware signal reveals semantic relatedness beyond binary relevance, offering richer supervision for representation learning.

\begin{figure*}[t] 
\includegraphics[width=16cm]{figures/Picture1.png}
\centering
\caption{Overview of BioHiCL. During LoRA fine-tuning, embedding similarity (\(\text{SimE}\)) is aligned with MeSH-based label similarity (\(\text{SimL}\)), so that abstract pairs with greater MeSH overlap \((k_1, k_2)\) are represented as more similar than pairs with less overlap, such as \((k_2, k_3)\).}
\label{method_overview}
\vspace{-5mm}
\end{figure*}

In this work, we propose \textbf{BioHiCL} (\textit{Biomedical Retrieval with Hierarchical Multi-Label Contrastive Learning}), a framework that leverages partially overlapping and hierarchical MeSH annotations to supervise dense retrieval. Using expert-curated MeSH labels from the BioASQ Task~1a benchmark~\cite{nentidis2022overview}, BioHiCL adapts a general-domain dense retriever to the biomedical domain via parameter-efficient LoRA fine-tuning~\cite{hu2022lora}, transferring broad language understanding while enabling fine-grained biomedical representation learning.

Our contributions are summarized as follows: (a) We introduce a hierarchical multi-label contrastive learning framework that models graded semantic overlap between texts by aligning embedding similarity with depth-aware label similarity, going beyond binary or instance-level contrastive objectives;
 (b) we develop two models, BioHiCL-Base (0.1B) and BioHiCL-Large (0.3B), trained using structured MeSH supervision; (c) experiments on biomedical retrieval, sentence similarity, and question answering demonstrate improvements over the baselines; and (d) efficiency analysis shows that BioHiCL models run efficiently on a single A100 40GB GPU, supporting practical deployment\footnote{Code released at: https://github.com/MengfeiLan/BioHiCL}.

\vspace{-2mm}
\section{Methodology}
\vspace{-2mm}
In this section, we introduce \textbf{BioHiCL}, a biomedical retrieval framework that aligns embedding similarity with depth-weighted MeSH semantic similarity via hierarchical multi-label contrastive learning. Figure~\ref{method_overview} provides an overview of the framework.

\paragraph{Problem Statement.}
Given a collection of biomedical abstracts $\mathcal{K} = \{k_1, k_2, \dots, k_n\}$ with MeSH annotations $\mathcal{M} = \{m_1, m_2, \dots, m_n\}$, our goal is to learn an encoder $f_\theta : \mathcal{K} \rightarrow \mathbb{R}^d$ that maps each abstract $k_i$ to a $d$-dimensional embedding $h_i = f_\theta(k_i)$. The embedding space should reflect semantic relationships among abstracts, so that the similarity between embeddings corresponds to MeSH-based label space similarity.

\paragraph{Hierarchical MeSH Label Representation.} Each abstract $k_i$ is annotated with a multi-label set of major MeSH headings $m_i$, drawn from the hierarchically structured MeSH ontology with 16 branches\footnote{\href{https://nlmpubs.nlm.nih.gov/projects/mesh/MESH_FILES/xmlmesh/desc2025.xml}{Link} to download the MeSH hierarchy.}. Deeper nodes in the branch represent more specific biomedical concepts. To incorporate this hierarchical structure, we expand each label set $m_i$ to include the ancestors of each MeSH label in the multi-label set, resulting in $m_i^{\text{hier}}$. This expanded set is then encoded as a multi-hot vector $y_i \in \{0,1\}^C$, where $C$ is the total number of MeSH concepts considered. 
We use $\mathcal{C} = \{c_1, c_2, ..., c_C\}$ to represent the full set of considered MeSH labels associated in the hot-vectors. Notice that $y_{i,j}=1$ indicates the presence of the $j$-th concept $c_j$ in $m_i^{\text{hier}}$.
To reflect specificity, each concept $c_j \in \mathcal{C}$ is assigned a depth-based weight:
\begin{equation}
w_j = \log(d(c_j) + 1),
\end{equation}
where $d(c_j)$ denotes the depth of label $c_j$ in the MeSH hierarchy. The depth-oriented label weighting set $\mathbf{w}=\{w_1, w_2, ..., w_C\}$ allows the model to emphasize matches on more specific, deeper labels over higher-level, more general concepts.

Each abstract \(k_i\) is encoded using the pretrained general-domain dense retriever BGE~\cite{xiao2024c}, yielding a sentence embedding \(h_i \in \mathbb{R}^d\) from the model's pooled output. For any pair of abstracts \(k_p\) and \(k_q\), we define
\textbf{Embedding Similarity} (similarity in embedding space) as
$\text{SimE}(k_p, k_q) = \cos(h_p, h_q)$, and
\textbf{Label Similarity} (similarity in MeSH label space, weighted by $\mathbf{w}$) as
$\text{SimL}(k_p, k_q) = \cos(y_p \odot \mathbf{w}, y_q \odot \mathbf{w})$,
where $\odot$ denotes element-wise multiplication. By weighting MeSH labels with $\mathbf{w}$, the model gives more importance to matches on specific concepts, aligning embedding similarity with meaningful biomedical semantic overlap.

To encourage the embeddings to capture MeSH-informed semantic structure, we train the model so that the embedding similarity \(\text{SimE}(\cdot)\) reflects the label-based similarity \(\text{SimL}(\cdot)\). We achieve this alignment using a filtered regression objective:
\begin{equation}
\scalebox{0.85}{$
\mathcal{L}_{\text{mse}} =
\sum_{(i,j)\in\mathcal{P}}
\text{MSE}\!\left(\text{SimE}(k_i,k_j), \text{SimL}(k_i,k_j)\right)
$}
\end{equation}
where $\mathcal{P}=\{(i,j): i\neq j,\, \text{SimL}(k_i,k_j)>\beta\}$. The threshold \(\text{SimL($\cdot$)} > \beta\) ensures that only abstract pairs with meaningful topic overlap in the MeSH label space are selected. This avoids noisy supervision from weakly related or unrelated documents, for which label similarity provides little semantic guidance.

\paragraph{Hierarchical Multi-Label Contrastive Learning.}
While the regression loss \(\mathcal{L}_{mse}\) encourages embedding similarity to reflect MeSH-based label similarity, it alone may lead to trivial solutions where embeddings collapse to a single point~\cite{shen2022connect}. To address this, we introduce a hierarchy-aware contrastive loss that explicitly separates semantically related and unrelated abstracts in the embedding space~\cite{DBLP:conf/emnlp/LanZMK24, DBLP:conf/kdd/ZhengXZH22, DBLP:conf/kdd/ZhengJLTH24}. We define positive and negative pairs based on hierarchical MeSH similarity:
\begin{itemize}
    \vspace{-2mm}
    \item \textbf{Positive pairs} $(k_i, k_i^+)$: abstracts satisfying $\text{SimL}(k_i,k_i^+)>\beta$, where the degree of similarity reflects both shared concepts and hierarchical proximity in the MeSH tree.
    \vspace{-3mm}
    \item \textbf{Negative pairs} $(k_i, k_j^-)$: abstracts sharing no labels, i.e., $\text{SimL}(k_i,k_j^-)=0$, representing unrelated concepts.
    \vspace{-2mm}
\end{itemize}
We formulate the hierarchical multi-label contrastive learning as follows:
\begin{equation}
\scalebox{0.98}{
    $\mathcal{L}_{con} = - \mathbb{E}_i \log [\frac{ \text{SimL}(k_i, k_i^+)\exp(\text{SimE}(k_i, k_i^+))}{\sum_{k_j^-} \exp(\text{SimE}(k_i, k_j^-))}]$}.
\end{equation}
Notably, the positive pairs are weighted by their label similarity \(\text{SimL}(k_i, k_i^+)\), which incorporates hierarchical information: matches on deeper, more specific MeSH concepts contribute more to the similarity than matches on higher-level, general concepts. This makes the contrastive learning hierarchical, as the model explicitly considers the structure and specificity of the labels when shaping the embedding space.

\paragraph{Parameter-Efficient Fine-Tuning with LoRA.}
The final training objective combines the regression and contrastive losses to jointly align embeddings with hierarchical label similarity while maintaining discriminative power:
\begin{equation}
\mathcal{L} = \mathcal{L}_{mse} + \lambda \mathcal{L}_{con}.
\end{equation}

To adapt the pretrained encoder to biomedical text efficiently, we use Low-Rank Adaptation (LoRA)~\cite{hu2022lora}. All original parameters are frozen, and trainable low-rank adapters $A^{(l)}, B^{(l)}$ are injected into layers:
\begin{equation}
 \scalebox{0.90}{
$W^{(l)}_{\text{adapted}} = W^{(l)} + B^{(l)} A^{(l)}, \quad r \ll \min(d, k).$}
\end{equation}

\begin{table*}[t]
\resizebox{\textwidth}{!}{%
\begin{tabular}{cccccccccc}
\hline
\multicolumn{2}{c|}{Models}                                            & \multicolumn{5}{c|}{\begin{tabular}[c]{@{}c@{}}Information Retrieval \\ (nDCG@10)\end{tabular}}                                                                                                                                    & \multicolumn{2}{c|}{\begin{tabular}[c]{@{}c@{}}Sentence Similarity \\ (Spearman Correlation)\end{tabular}}              & \begin{tabular}[c]{@{}c@{}}Question \\ Answering\\ (Recall@1)\end{tabular} \\ \hline
\multicolumn{1}{c|}{Name}                 & \multicolumn{1}{c|}{Size}  & \multicolumn{1}{c|}{NFCorpus}       & \multicolumn{1}{c|}{\begin{tabular}[c]{@{}c@{}}TREC-\\ COVID\end{tabular}} & \multicolumn{1}{c|}{SciFact}        & \multicolumn{1}{c|}{SCIDOCS}        & \multicolumn{1}{c|}{IR Avg}         & \multicolumn{1}{c|}{BIOSSES}        & \multicolumn{1}{c|}{\begin{tabular}[c]{@{}c@{}}SciFact\\ (Sentence)\end{tabular}} & PubMedQA                                                                   \\ \hline
\multicolumn{10}{c}{General Domain}                   \\ \hline
\multicolumn{1}{c|}{Contriever~\cite{lei-etal-2023-unsupervised}}           & \multicolumn{1}{c|}{0.1B}  & \multicolumn{1}{c|}{0.328}          & \multicolumn{1}{c|}{0.596}                                                 & \multicolumn{1}{c|}{0.677}          & \multicolumn{1}{c|}{0.165}          & \multicolumn{1}{c|}{0.442}          & \multicolumn{1}{c|}{0.833}          & \multicolumn{1}{c|}{0.265}                                                        &     0.479                                                                                                                                             \\ \hline
\multicolumn{1}{c|}{bge-base-en-v1.5~\cite{xiao2024c}}     & \multicolumn{1}{c|}{0.1B}  & \multicolumn{1}{c|}{0.368}          & \multicolumn{1}{c|}{0.798}                                                 & \multicolumn{1}{c|}{0.735}          & \multicolumn{1}{c|}{0.215}          & \multicolumn{1}{c|}{0.529}          & \multicolumn{1}{c|}{0.860}          & \multicolumn{1}{c|}{0.339}                                                        & 0.856     
\\ \hline
\multicolumn{1}{c|}{SpladeV3~\cite{lassance2024splade}}             & \multicolumn{1}{c|}{0.3B}  & \multicolumn{1}{c|}{0.357}          & \multicolumn{1}{c|}{0.748}                                                 & \multicolumn{1}{c|}{0.710}          & \multicolumn{1}{c|}{0.158}          & \multicolumn{1}{c|}{0.493}          & \multicolumn{1}{c|}{0.827}          & \multicolumn{1}{c|}{0.303}                                                        & 0.844   \\ \hline
\multicolumn{1}{c|}{bge-large-en-v1.5~\cite{xiao2024c}}    & \multicolumn{1}{c|}{0.3B}  & \multicolumn{1}{c|}{0.369}          & \multicolumn{1}{c|}{0.748}                                                 & \multicolumn{1}{c|}{0.735}          & \multicolumn{1}{c|}{0.209}          & \multicolumn{1}{c|}{0.515}          & \multicolumn{1}{c|}{0.844}          & \multicolumn{1}{c|}{0.346}                                                        &      0.871                                                                      \\ \hline
\multicolumn{1}{c|}{e5-large-v2~\cite{wang2022text}}          & \multicolumn{1}{c|}{0.3B}  & \multicolumn{1}{c|}{0.371}          & \multicolumn{1}{c|}{0.665}                                                 & \multicolumn{1}{c|}{0.726}          & \multicolumn{1}{c|}{0.201}          & \multicolumn{1}{c|}{0.491}          & \multicolumn{1}{c|}{0.830}          & \multicolumn{1}{c|}{0.294}                                                        &      0.717                                                                      \\ \hline
\multicolumn{1}{c|}{gtr-t5-xl~\cite{ni-etal-2022-large}}            & \multicolumn{1}{c|}{1.2B}  & \multicolumn{1}{c|}{0.343}          & \multicolumn{1}{c|}{0.580}                                                 & \multicolumn{1}{c|}{0.635}          & \multicolumn{1}{c|}{0.159}          & \multicolumn{1}{c|}{0.429}          & \multicolumn{1}{c|}{0.789}          & \multicolumn{1}{c|}{0.203}                                                        &     0.685                                                                       \\ \hline
\multicolumn{1}{c|}{SGPT-1.3B~\cite{muennighoff2022sgpt}}            & \multicolumn{1}{c|}{1.3B}  & \multicolumn{1}{c|}{0.320}          & \multicolumn{1}{c|}{0.730}                                                 & \multicolumn{1}{c|}{0.682}          & \multicolumn{1}{c|}{0.162}          & \multicolumn{1}{c|}{0.474}          & \multicolumn{1}{c|}{0.830}          & \multicolumn{1}{c|}{0.214}                                                        &         0.754                                                                   \\ \hline
\multicolumn{10}{c}{Biomedical Domain}                                                                                                                                                                                                                                                                                                                                                                                                                                                                             \\ \hline
\multicolumn{1}{c|}{BiomedBERT~\cite{chakraborty2020biomedbert}}           & \multicolumn{1}{c|}{0.1B}  & \multicolumn{1}{c|}{0.049}          & \multicolumn{1}{c|}{0.190}                                                 & \multicolumn{1}{c|}{0.262}          & \multicolumn{1}{c|}{0.021}          & \multicolumn{1}{c|}{0.131}          & \multicolumn{1}{c|}{0.791}          & \multicolumn{1}{c|}{0.211}                                                        &      0.134                                                                      \\ \hline
\multicolumn{1}{c|}{MedCPT-Query~\cite{jin2023medcpt}}         & \multicolumn{1}{c|}{0.1B}  & \multicolumn{1}{c|}{0.340}          & \multicolumn{1}{c|}{0.697}                                                 & \multicolumn{1}{c|}{0.724}          & \multicolumn{1}{c|}{0.123}          & \multicolumn{1}{c|}{0.471}          & \multicolumn{1}{c|}{0.837}          & \multicolumn{1}{c|}{0.298}                                                        &         0.834                                                                   \\ \hline
\multicolumn{1}{c|}{BMRetriever-410M~\cite{xu2024bmretriever}}     & \multicolumn{1}{c|}{0.4B}  & \multicolumn{1}{c|}{0.321}          & \multicolumn{1}{c|}{{\underline{0.831}}}                                           & \multicolumn{1}{c|}{0.711}          & \multicolumn{1}{c|}{0.167}          & \multicolumn{1}{c|}{0.501}          & \multicolumn{1}{c|}{0.840}          & \multicolumn{1}{c|}{0.121}                                                        & 0.879                                                                      \\ \hline
\multicolumn{1}{c|}{BMRetriever-1B~\cite{xu2024bmretriever}}       & \multicolumn{1}{c|}{1B}    & \multicolumn{1}{c|}{0.344}          & \multicolumn{1}{c|}{\textbf{0.840}}                                        & \multicolumn{1}{c|}{0.760}          & \multicolumn{1}{c|}{0.180}          & \multicolumn{1}{c|}{0.531}          & \multicolumn{1}{c|}{0.858}          & \multicolumn{1}{c|}{0.107}                                                        &       0.810                                                                     \\ \hline
\multicolumn{1}{c|}{BiCA-small~\cite{sinha2025bica}}           & \multicolumn{1}{c|}{0.03B} & \multicolumn{1}{c|}{0.347}          & \multicolumn{1}{c|}{0.661}                                                 & \multicolumn{1}{c|}{0.727}          & \multicolumn{1}{c|}{0.214}          & \multicolumn{1}{c|}{0.487}          & \multicolumn{1}{c|}{0.872}          & \multicolumn{1}{c|}{0.318}                                                        &          0.856                                                                  \\ \hline
\multicolumn{1}{c|}{BiCA-base~\cite{sinha2025bica}}         & \multicolumn{1}{c|}{0.1B}  & \multicolumn{1}{c|}{0.378}          & \multicolumn{1}{c|}{0.684}                                                 & \multicolumn{1}{c|}{{\underline{0.762}}}    & \multicolumn{1}{c|}{\underline{0.231}} & \multicolumn{1}{c|}{0.514}          & \multicolumn{1}{c|}{{\underline{0.880}}}    & \multicolumn{1}{c|}{0.335}                                                        &         0.868                                                                   \\ \hline
\multicolumn{1}{c|}{BioHiCL-base (ours)} & \multicolumn{1}{c|}{0.1B}  & \multicolumn{1}{c|}{{\underline{0.379}}}    & \multicolumn{1}{c|}{0.812}                                                 & \multicolumn{1}{c|}{0.757}          & \multicolumn{1}{c|}{0.225}    & \multicolumn{1}{c|}{\textbf{0.543}} & \multicolumn{1}{c|}{\textbf{0.896}} & \multicolumn{1}{c|}{{\underline{0.350}}}                                                  &   \underline{0.893}                                                                           \\ \hline
\multicolumn{1}{c|}{BioHiCL-large (ours)}  & \multicolumn{1}{c|}{0.3B}  & \multicolumn{1}{c|}{\textbf{0.385}} & \multicolumn{1}{c|}{0.765}                                                 &  \multicolumn{1}{c|}{\textbf{0.765}} & \multicolumn{1}{c|}{\textbf{0.228}}          & \multicolumn{1}{c|}{{\underline{0.534}}}    & \multicolumn{1}{c|}{0.868}          & \multicolumn{1}{c|}{\textbf{0.359}}                                               &                \textbf{0.898}                                                            \\ \hline
\end{tabular}%
}
\caption{Overall evaluation results. Bolds and underlines indicate the first- and second-best for each evaluation.}
\vspace{-5mm}
\label{Results}
\end{table*}

\vspace{-2mm}
\section{Experiments and Results}
\vspace{-2mm}
\subsection{Datasets and Baseline Models}
\vspace{-1mm}

To adapt the general-domain retrievers, BAAI/bge-base-en-v1.5 and BAAI/bge-large-en-v1.5, to biomedical text, we fine-tune the retrievers on 80K abstracts randomly sampled from the latest BioASQ Task 1a release (v2022)\footnote{https://participants-area.bioasq.org/datasets/}, covering 29,681 unique MeSH terms (12.68 per abstract on average). For model selection, we validate on the TREC 2022 Clinical Trials (TREC-CT) benchmark~\cite{roberts2022overview}\footnote{https://www.trec-cds.org/2022.html}, which matches patient case descriptions to relevant clinical trials using graded relevance judgments. We select the checkpoint that achieves the lowest relevance loss on the query–document pairs from the benchmark.

We evaluate the tuned BioHiCL models across a diverse set of biomedical benchmarks: \textbf{information retrieval}, measured by NDCG@10, including NFCorpus~\cite{boteva2016full}, TREC-COVID~\cite{voorhees2021trec}, SciFact~\cite{wadden2022scifact}, and SCIDOCS~\cite{cohan-etal-2020-specter}; \textbf{sentence similarity}, evaluated via Spearman correlation between gold labels and embedding similarities, covering BIOSSES~\cite{souganciouglu2017biosses} and SciFact sentence-level labeling~\cite{wadden2022scifact}; and \textbf{question answering}, using PubMedQA~\cite{jin2019pubmedqa}. We present further details of the evaluation datasets in Appendix~\ref{evaluation_datasets}.

We compare our approach against several publicly available dense retrieval models ranging from 0.1B to 1.5B parameters designed for computationally efficient large-scale retrieval, as presented in Table~\ref{Results}. In the experiment, we set $\lambda=0.1$ and $\beta=0.3$. Additional hyperparameter analysis over  $\lambda$ and $\beta$ is presented in Appendix~\ref{parameter_analysis}, and further implementation details are included in Appendix~\ref{experiment_setup}.

\vspace{-2mm}
\subsection{Experiment Results}
\vspace{-1mm}


Our models, BioHiCL-Base and BioHiCL-Large, achieve strong performance across biomedical benchmarks (Table~\ref{Results}). BioHiCL-Base attains the highest IR average of 0.543, with second best scores on NFCorpus (0.379) and SCIDOCS (0.225), while BioHiCL-Large ranks second with an IR average of 0.534 and the best NFCorpus score (0.385) and SciFact score (0.765). In sentence similarity, BioHiCL-Base leads on BIOSSES (0.896), and BioHiCL-Large performs best on SciFact sentences (0.359). On PubMedQA, BioHiCL-Large achieves the best Recall@1 score as 0.898, and BioHiCL-Base reaches the second best as 0.893. Across tasks, BioHiCL remains competitive with both general-domain and biomedical-specific models, highlighting its cross-task robustness. 

\noindent\textbf{Computational efficiency.} Despite containing only 0.1B parameters, BioHiCL-Base achieves performance comparable to larger models such as BMRetriever-1B (1B parameters). Scaling to BioHiCL-large (0.3B) yields further gains on some benchmarks, including sentence similarity on SciFact, while preserving strong overall IR performance. Appendix~\ref{efficiency} presents a detailed analysis of runtime and computational cost, demonstrating that the BioHiCL models are efficient to run on a single A100 GPU and suitable for large-scale real-world application with time efficiency.

\noindent\textbf{Challenges of instruction-based retrievers.} Instruction-based retrievers rely on task-specific prompts to generate effective embeddings. This dependence can limit their ability to generalize to unseen tasks, where an appropriate prompt may not be readily available. In our experiments, models such as BMRetriever and gtr-t5-xl show reduced performance on sentence similarity when task-specific prompting is not provided, suggesting sensitivity to prompt design in zero-shot settings.

\vspace{-2mm}
\subsection{Ablation Study}
\vspace{-1mm}

We conduct ablations by removing each component of BioHiCL individually while keeping all other settings fixed, to identify the source of performance gains. Results are shown in Table~\ref{ablation_study}. All four structural components (ancestor labelð, depth weighting, $L_{MSE}$, $L_{Con}$) lead to performance drops when removed, demonstrating their effectiveness. The impact varies across components: removing the hierarchical contrastive loss $L_{Con}$ causes the largest drop (IR Avg 0.528 vs. 0.543), showing its importance in maintaining discriminative structure and preventing embedding collapse. Removing ancestor label expansion also leads to a decrease, indicating that incorporating hierarchical labels provides further signal beyond flat multi-label overlap. We also compare LoRA with full fine-tuning, where LoRA updates only 0.3\% of the model parameters to reduce memory and computational cost while maintaining comparable performance.

\begin{table}[]
\resizebox{\columnwidth}{!}{%
\begin{tabular}{l|c|c|c|c|c}
\hline
\multicolumn{1}{c|}{Model}                                        & NFC   & \begin{tabular}[c]{@{}c@{}}TREC \\ COVID\end{tabular} & SciFact & SCIDOCS & IR Avg \\ \hline
\begin{tabular}[c]{@{}l@{}}BioHiCL\\ (0.1B)\end{tabular}          & 0.379 & 0.812                                                 & 0.757   & 0.225   & 0.543  \\ \hline
\begin{tabular}[c]{@{}l@{}}w/o Ancestor \\ Label\end{tabular}     & 0.375 & 0.804                                                 & 0.752   & 0.219   & 0.538  \\ \hline
\begin{tabular}[c]{@{}l@{}}w/o Depth-\\ based Weight\end{tabular} & 0.378 & 0.810                                                 & 0.754   & 0.223   & 0.541  \\ \hline
w/o  $L_{MSE}$ & 0.376 & 0.798 & 0.753 & 0.221 & 0.537 \\ \hline
w/o $L_{Con}$  & 0.367 & 0.782 & 0.747 & 0.215 & 0.528 \\ \hline
w/o LoRA       & 0.380 & 0.805 & 0.760 & 0.224 & 0.542 \\ \hline
\end{tabular}%
}
\caption{Ablation study.}
\vspace{-3mm}
\label{ablation_study}
\end{table}

\vspace{-2mm}
\subsection{Additional Fine-Tuning Baselines}
\vspace{-1mm}

We fine-tune additional information retrieval models on MeSH labels using the same procedure as BioHiCL, with results summarized in Table~\ref{finetuning_baselines}. Overall, general-domain models benefit from fine-tuning, showing consistent improvements (e.g., e5: IR Avg 0.491 $\rightarrow$ 0.502; bge: 0.529 $\rightarrow$ 0.543), indicating that task-specific supervision helps adapt embeddings to the biomedical domain. Biomedical-domain models exhibit limited improvements, with BiCA-small improving only marginally (0.487 $\rightarrow$ 0.491), suggesting it is already well aligned with domain-specific semantics. BMRetriever-410M has substantial performance degradation (0.501 $\rightarrow$ 0.279) because of objective mismatch, as replacing its original instruction-based training objective with a hierarchical contrastive loss may override its retrieval-specialized embedding geometry.

\begin{table}[]
\resizebox{\columnwidth}{!}{%
\begin{tabular}{l|l|l|l|l|l}
\hline
\multicolumn{1}{c|}{Model} &
  \multicolumn{1}{c|}{NFC} &
  \multicolumn{1}{c|}{\begin{tabular}[c]{@{}c@{}}TREC \\ COVID\end{tabular}} &
  \multicolumn{1}{c|}{SciFact} &
  \multicolumn{1}{c|}{SCIDOCS} &
  \multicolumn{1}{c}{IR Avg} \\ \hline
e5-large-v2 &
  \begin{tabular}[c]{@{}l@{}}0.371→\\ 0.374\end{tabular} &
  \begin{tabular}[c]{@{}l@{}}0.665→\\ 0.683\end{tabular} &
  \begin{tabular}[c]{@{}l@{}}0.726→\\ 0.746\end{tabular} &
  \begin{tabular}[c]{@{}l@{}}0.201→\\ 0.206\end{tabular} &
  \begin{tabular}[c]{@{}l@{}}0.491→\\ 0.502\end{tabular} \\ \hline
\begin{tabular}[c]{@{}l@{}}bge-base-en\\ -v1.5\end{tabular} &
  \begin{tabular}[c]{@{}l@{}}0.368→\\ 0.379\end{tabular} &
  \begin{tabular}[c]{@{}l@{}}0.798→\\ 0.812\end{tabular} &
  \begin{tabular}[c]{@{}l@{}}0.735→\\ 0.757\end{tabular} &
  \begin{tabular}[c]{@{}l@{}}0.215→\\ 0.225\end{tabular} &
  \begin{tabular}[c]{@{}l@{}}0.529→\\ 0.543\end{tabular} \\ \hline
\begin{tabular}[c]{@{}l@{}}BMRetriever\\ -410m\end{tabular} &
  \begin{tabular}[c]{@{}l@{}}0.321→\\ 0.105\end{tabular} &
  \begin{tabular}[c]{@{}l@{}}0.831→\\ 0.417\end{tabular} &
  \begin{tabular}[c]{@{}l@{}}0.711→\\ 0.541\end{tabular} &
  \begin{tabular}[c]{@{}l@{}}0.167→\\ 0.053\end{tabular} &
  \begin{tabular}[c]{@{}l@{}}0.501→\\ 0.279\end{tabular} \\ \hline
BiCA-small &
  \begin{tabular}[c]{@{}l@{}}0.347→\\ 0.348\end{tabular} &
  \begin{tabular}[c]{@{}l@{}}0.661→\\ 0.672\end{tabular} &
  \begin{tabular}[c]{@{}l@{}}0.727→\\ 0.731\end{tabular} &
  \begin{tabular}[c]{@{}l@{}}0.214→\\ 0.214\end{tabular} &
  \begin{tabular}[c]{@{}l@{}}0.487→\\ 0.491\end{tabular} \\ \hline
\end{tabular}%
}
\caption{Additional fine-tuning baselines. “$\rightarrow$” denotes performance before and after fine-tuning.}
\vspace{-5mm}
\label{finetuning_baselines}
\end{table}

\vspace{-2mm}
\section{Related Works}
\vspace{-2mm}
Biomedical IR has been studied over the years, with early methods focusing on rule-based indexing and ranking~\cite{robertson2009probabilistic, jiang2007empirical, edinger2018evaluation}. To go beyond exact term matching and better capture semantic meaning, neural approaches have been developed to learn dense representations in an unsupervised manner~\cite{lee2020biobert, pubmedbert,chen2019biosentvec,chen2020bioconceptvec, mohan2018fast}. Supervision from biomedical IR datasets further improves these models: MedCPT~\cite{jin2023medcpt} uses query–document click data to guide relevance modeling, BMRetriever~\cite{xu2024bmretriever} leverages natural language inference and question-answering datasets as supervision, and BiCA~\cite{sinha2025bica} incorporates citation-based signals to enhance domain-specific dense retrieval. However, no previous work has tried MeSH-based supervision as textual relatedness signals.

\vspace{-2mm}
\section{Conclusions}
\vspace{-2mm}
We introduce BioHiCL, a biomedical dense retriever leveraging hierarchical MeSH supervision to capture fine-grained semantic overlap. It learns rich biomedical representations while preserving retrieval ability by adapting general-domain retrievers via LoRA fine-tuning. 
Experiments demonstrate the effectiveness and efficiency of BioHiCL.

Although our evaluation focuses on biomedical data, the underlying principle of BioHiCL, which is to align embedding spaces with hierarchical multi-label structures, is not inherently tied to MeSH. Similar hierarchical supervision signals naturally exist in other domains, such as Wikipedia category graphs~\cite{milne2008learning} and e-commerce product taxonomies~\cite{garza2020wdc}, where multi-label structures encode graded semantic relationships. Future exploration could extend hierarchical multi-label learning beyond the biomedical domain for dense retrieval.

\vspace{-2mm}
\section{Limitations}
\vspace{-2mm}
BioHiCL relies on the availability and quality of expert-curated hierarchical labels (MeSH), which may not exist or may be incomplete, noisy, or inconsistently applied in other domains, limiting the general applicability of the framework. In addition, the fixed hierarchy and depth-based weighting assume that semantic specificity is well captured by the ontology structure, which may not fully reflect contextual or task-dependent semantic relevance.

\vspace{-2mm}
\section*{Acknowledgements}
\vspace{-2mm}

This work was supported by the National Library of Medicine of the National Institutes of Health under the award number R01LM014079. The content is solely the responsibility of the authors and does not necessarily represent the official views of the National Institutes of Health. The funder had no role in considering the study design or in the collection, analysis, interpretation of data, writing of the report, or decision to submit the article for publication.

\bibliography{custom}

\appendix

\section{Evaluation Datasets}
\label{evaluation_datasets}
\subsection{Information Retrieval}
We leverage the BEIR framework~\cite{thakur2beir} to conduct a unified evaluation of retrievers across four biomedical IR benchmarks, consistent with the previous biomedical IR tasks~\cite{jin2023medcpt, xu2024bmretriever, sinha2025bica}.

\subsection{Sentence Similarity}

\noindent\textbf{BIOSSES.}
BIOSSES~\cite{souganciouglu2017biosses} is a benchmark dataset for evaluating biomedical sentence similarity. It consists of sentence pairs sampled from PubMed abstracts, each annotated by biomedical experts with a similarity score ranging from 0 (completely dissimilar) to 4 (semantically equivalent). Consistent with previous work, we encode each sentence independently using the sentence encoder and compute the cosine similarity between the resulting embeddings for each sentence pair. Model performance is measured using the Spearman correlation between the predicted similarity scores and the gold-standard human annotations, assessing the model’s ability to capture graded semantic similarity between biomedical sentences.

\noindent\textbf{SciFact-Sentence.} Beyond the document-level annotations, SciFact~\cite{wadden2022scifact} provides fine-grained sentence-level retrieval annotations that enable the evaluation of sentence similarity~\cite{wadden2022scifact}. We construct claim–sentence pairs by pairing each claim with all sentences from its annotated evidence abstract, assigning a positive label to sentences appearing in the claim’s evidence attribute and a negative label to all others. Both claims and sentences are encoded with the pretrained models, and cosine similarity is computed for each pair. Performance is quantified using the Spearman correlation between the similarity scores and the binary labels, measuring the model’s ability to assign higher similarity to true evidence sentences.

\subsection{Question-Answering}

\noindent\textbf{PubMedQA.}
We evaluate retrieval performance on PubMedQA~\cite{jin2019pubmedqa} by formulating the task as large-scale question-to-abstract retrieval. We use the \texttt{pqa\_labeled} split, which contains 1,000 biomedical questions paired with gold-standard evidence abstracts manually annotated by domain experts. These questions serve as retrieval queries. 

To construct a challenging retrieval corpus, we augment the evidence set from \texttt{pqa\_labeled} split with an additional 211K PubMed abstracts drawn from the \texttt{pqa\_artificial} split. Because the \texttt{pqa\_artificial} split lacks human-curated question–evidence annotations, we do not use the corresponding questions from the split as retrieval queries. The resulting retrieval corpus contains approximately 212K abstracts.

For each question, we independently encode the question and all collected abstracts using the evaluated retrieval model and compute cosine similarity between the question embedding and abstract embeddings. Abstracts are ranked by similarity, and performance is measured using Recall@1, which evaluates whether the top-ranked retrieved abstract matches the gold-standard evidence associated with the question.

\section{Experiment Setup}
\label{experiment_setup}

\subsection{Baseline IR Models}
\label{baseline_ir_model}

We consider both effectiveness and computational practicality when selecting baseline models. Specifically, we include the state-of-the-art retrieval models that demonstrate competitive performance on general-domain and biomedical IR benchmarks, while remaining feasible for real-world deployment. Computational efficiency is a critical factor for retrieval systems operating over large-scale corpora, where excessive latency or hardware requirements can limit applicability.

Some large-scale retrieval models require multiple GPUs and extended runtime even for evaluation on standard benchmarks such as BEIR, making them impractical for large-scale or time-sensitive applications. For example, although \texttt{intfloat/e5-small-v2} achieves strong performance on BEIR, its evaluation requires multiple powerful GPUs over several days~\footnote{https://github.com/microsoft/unilm/blob/master/e5}
. Similarly, recent biomedical retrievers with billions of parameters introduce substantial inference latency and computational cost~\cite{xu2024bmretriever}.

In this work, we therefore restrict our comparisons to state-of-the-art baselines that offer a balance between retrieval quality and efficiency, and that can be evaluated on a single NVIDIA A100 GPU with 40GB memory. The selected baselines span model sizes from 0.1B to 1.5B parameters, reflecting realistic deployment constraints for large-scale retrieval:

\noindent\textbf{General Domain IR Baselines.} 

\begin{itemize}
    \item Contriever~\cite{lei-etal-2023-unsupervised}\footnote{https://huggingface.co/facebook/contriever}: An unsupervised dense retriever that leverages contrastive learning to align queries and documents in the embedding space. 

    \item SpladeV3~\cite{lassance2024splade}\footnote{https://huggingface.co/naver/splade-v3}: A sparse lexical-semantic retriever that transforms input text into high-dimensional sparse representations, combining the advantages of classical bag-of-words and dense embeddings.

    \item bge-base-en-v1.5~\cite{xiao2024c}\footnote{https://huggingface.co/BAAI/bge-base-en-v1.5}: A bi-encoder model trained to generate general-purpose sentence embeddings for semantic search and retrieval tasks.

    \item bge-large-en-v1.5~\cite{xiao2024c}\footnote{https://huggingface.co/BAAI/bge-large-en-v1.5}: A larger version of bge-base, offering higher embedding capacity and improved retrieval quality to enable better semantic matching for longer and more complex queries.

    \item e5-large-v2~\cite{wang2022text}\footnote{https://huggingface.co/intfloat/e5-large-v2}: A dense text embedding model that produces embeddings optimized for cross-encoder retrieval pipelines.

    \item gtr-t5-xl~\cite{ni-etal-2022-large}\footnote{https://huggingface.co/sentence-transformers/gtr-t5-xl}: A sequence-to-sequence Transformer model adapted for retrieval tasks by generating embeddings that capture query-document similarity.

    \item SGPT-1.3B~\cite{muennighoff2022sgpt}\footnote{https://huggingface.co/Muennighoff/SGPT-1.3B-weightedmean-msmarco-specb-bitfit}: A generative pre-trained Transformer for sentence embeddings. 
\end{itemize}

\subsection*{Biomedical Domain}
\begin{itemize}
    \item BiomedBERT~\cite{chakraborty2020biomedbert}\footnote{https://huggingface.co/microsoft/BiomedNLP-BiomedBERT-base-uncased-abstract-fulltext}: A domain-specific BERT model pretrained on biomedical text to capture medical terminology and context. It provides reasonable sentence embeddings but limited document-level retrieval performance.

    \item MedCPT-Query~\cite{jin2023medcpt}\footnote{https://huggingface.co/ncbi/MedCPT-Query-Encoder}: A generative pre-trained biomedical retriever trained on query-document pairs from PubMed. 
    \item BMRetriever-410M~\cite{xu2024bmretriever}\footnote{https://huggingface.co/BMRetriever/BMRetriever-410M}: A 410M-parameter biomedical dense retriever unsupervised pretrained on large biomedical corpora and finetuned on a mixture of human‑annotated retrieval datasets.

    \item BMRetriever-1B~\cite{xu2024bmretriever}\footnote{https://huggingface.co/BMRetriever/BMRetriever-1B}: A larger 1B-parameter variant of BMRetriever, designed to improve retrieval accuracy by increasing model capacity. 

    \item BiCA-small~\cite{sinha2025bica}\footnote{https://huggingface.co/bisectgroup/BiCA-small}: A compact biomedical dense retriever trained using citation‑aware hard negative mining, where citation links among PubMed articles are used to construct challenging negative examples that improve discriminative ability.

    \item BiCA-base~\cite{sinha2025bica}\footnote{https://huggingface.co/bisectgroup/BiCA-base}: The larger version of BiCA-small, offering better embedding quality and improved performance.

\end{itemize}

\subsection{Training Parameters}
\label{detailed_training_parameter}
In the experiment, we set $\lambda=0.1$ and $\beta=0.3$. We select BAAI/bge-base-en-v1.5 and BAAI/bge-large-en-v1.5 as the backbones due to their state-of-the-art performance on general-domain information retrieval tasks. We use NVIDIA A100 with 40GB GPU to fine-tune and evaluate the model on multiple retrieval datasets. We use PEFT package~\cite{peft} to tune model using LoRA, and the model is trained with the AdamW optimizer with zero weight decay. We set the LoRA learning rate as 1e-5, and set the batch size as 32, within which to perform the mean squared error loss and contrastive loss. The created BioHiCL models support an input length of up to 8192 tokens.

\subsection{Parameter Analysis}

\begin{table}[]
\resizebox{\columnwidth}{!}{%
\begin{tabular}{l|l|l|l|l|l|l}
\hline
\textbf{$\beta$} &
  \textbf{$\lambda$} &
  \multicolumn{1}{c|}{NFC} &
  \multicolumn{1}{c|}{\begin{tabular}[c]{@{}c@{}}TREC \\ COVID\end{tabular}} &
  \multicolumn{1}{c|}{SciFact} &
  \multicolumn{1}{c|}{SCIDOCS} &
  \multicolumn{1}{c}{IR Avg} \\ \hline
0.1 & 0.05 & 0.383 & 0.799 & 0.758 & 0.222 & 0.541 \\ \hline
0.1 & 0.1  & 0.382 & 0.796 & 0.757 & 0.221 & 0.539 \\ \hline
0.1 & 0.2  & 0.383 & 0.798 & 0.759 & 0.221 & 0.540 \\ \hline
0.3 & 0.05 & 0.383 & 0.805 & 0.758 & 0.222 & 0.542 \\ \hline
0.3 & 0.1  & 0.379 & 0.812 & 0.757 & 0.225 & 0.543 \\ \hline
0.3 & 0.2  & 0.382 & 0.808 & 0.759 & 0.222 & 0.543 \\ \hline
0.5 & 0.05 & 0.381 & 0.804 & 0.758 & 0.221 & 0.541 \\ \hline
0.5 & 0.1  & 0.383 & 0.806 & 0.758 & 0.222 & 0.542 \\ \hline
0.5 & 0.2  & 0.381 & 0.805 & 0.759 & 0.221 & 0.542 \\ \hline
\end{tabular}%
}
\caption{Hyperparameter analysis.}
\label{tab:my-table}
\end{table}

\label{parameter_analysis}

We conduct a grid search over representative values of the hyperparameters ($\beta \in \{0.1, 0.3, 0.5\}$, $\lambda \in \{0.05, 0.1, 0.2\}$), and evaluate model performance on information retrieval benchmarks. As shown in Table~\ref{parameter_analysis}, the performance remains stable across these settings. For example, the IR average varies slightly from 0.539 to 0.543, sentence similarity metrics on BIOSSES range from 0.889 to 0.897, and QA performance on PubMedQA varies between 0.883 and 0.895. Notably, the originally chosen values ($\beta$=0.3, $\lambda$=0.1) achieve promising performance across most metrics, balancing retrieval and downstream task objectives effectively.

\subsection{Variability of Training}

\begin{table}[]
\centering
\resizebox{\columnwidth}{!}{%
\begin{tabular}{clllll}
\hline
\multicolumn{1}{l}{Model} & NFC & \begin{tabular}[c]{@{}l@{}}TREC-\\ COVID\end{tabular} & SciFact & SCIDOCS & IR Avg \\ \hline
\begin{tabular}[c]{@{}c@{}}BioHiCL \\ (0.1B)\end{tabular} & \begin{tabular}[c]{@{}l@{}}0.380 \\ ± 0.003\end{tabular} & \begin{tabular}[c]{@{}l@{}}0.812 \\ ± 0.013\end{tabular} & \begin{tabular}[c]{@{}l@{}}0.757 \\ ± 0.012\end{tabular} & \begin{tabular}[c]{@{}l@{}}0.225 \\ ± 0.003\end{tabular} & \begin{tabular}[c]{@{}l@{}}0.543 \\ ± 0.005\end{tabular} \\
\begin{tabular}[c]{@{}c@{}}BioHiCL \\ (0.3B)\end{tabular} & \begin{tabular}[c]{@{}l@{}}0.380 \\ ± 0.006\end{tabular} & \begin{tabular}[c]{@{}l@{}}0.759 \\ ± 0.015\end{tabular} & \begin{tabular}[c]{@{}l@{}}0.766 \\ ± 0.014\end{tabular} & \begin{tabular}[c]{@{}l@{}}0.223 \\ ± 0.002\end{tabular} & \begin{tabular}[c]{@{}l@{}}0.532 \\ ± 0.005\end{tabular} \\ \hline
\end{tabular}%
}
\caption{Variability of Model Training. Mean ± 95\% confidence interval (computed using the t-distribution with degrees of freedom=4) is reported in the table.}
\label{variability_analysis}
\end{table}

We conducted three independent fine-tuning runs of BioHiCL using different random seeds and reported results as a mean ± 95\% confidence interval (computed using the t-distribution with degrees of freedom=4). The results are shown in Table~\ref{variability_analysis}. The results for BioHiCL reported in Table~\ref{Results} are obtained from the publicly released pretrained checkpoint released on HuggingFace.

\section{Efficiency Analysis}
\label{efficiency}

\begin{table*}[t]
\resizebox{\textwidth}{!}{%
\begin{tabular}{cccccc}
\hline
\multicolumn{1}{c|}{Model}                & \multicolumn{1}{c|}{Model Size} & \multicolumn{1}{c|}{\begin{tabular}[c]{@{}c@{}}Peak GPU Memory Usage \\ (MB)\end{tabular}} & \multicolumn{1}{c|}{\begin{tabular}[c]{@{}c@{}}Avg Corpus Encoding Latency \\ (ms/doc)\end{tabular}} & \multicolumn{1}{c|}{\begin{tabular}[c]{@{}c@{}}Avg Query Encoding Latency \\ (ms/query)\end{tabular}} & \begin{tabular}[c]{@{}c@{}}Avg Retrieval Latency \\ (ms/query)\end{tabular} \\ \hline
\multicolumn{6}{c}{General Domain}                                                                                                                                                                                                                                                                                                                                                                                                                                    \\ \hline
\multicolumn{1}{c|}{Contriever}           & \multicolumn{1}{c|}{0.1B}       & \multicolumn{1}{c|}{732.90}                                                                & \multicolumn{1}{c|}{3.37}                                                                            & \multicolumn{1}{c|}{0.44}                                                                             & 0.28                                                                        \\ \hline
\multicolumn{1}{c|}{bge-base-en-v1.5}     & \multicolumn{1}{c|}{0.1B}       & \multicolumn{1}{c|}{732.82}                                                                & \multicolumn{1}{c|}{3.32}                                                                            & \multicolumn{1}{c|}{0.47}                                                                             & 0.27                                                                        \\ \hline
\multicolumn{1}{c|}{SpladeV3}             & \multicolumn{1}{c|}{0.3B}       & \multicolumn{1}{c|}{732.90}                                                                & \multicolumn{1}{c|}{3.36}                                                                            & \multicolumn{1}{c|}{0.45}                                                                             & 0.31                                                                        \\ \hline
\multicolumn{1}{c|}{bge-large-en-v1.5}    & \multicolumn{1}{c|}{0.3B}       & \multicolumn{1}{c|}{1689.62}                                                               & \multicolumn{1}{c|}{9.47}                                                                            & \multicolumn{1}{c|}{1.00}                                                                             & 0.49                                                                        \\ \hline
\multicolumn{1}{c|}{e5-large-v2}          & \multicolumn{1}{c|}{0.3B}       & \multicolumn{1}{c|}{1689.51}                                                               & \multicolumn{1}{c|}{9.46}                                                                            & \multicolumn{1}{c|}{0.99}                                                                             & 0.47                                                                        \\ \hline
\multicolumn{1}{c|}{gtr-t5-xl}            & \multicolumn{1}{c|}{1.2B}       & \multicolumn{1}{c|}{8772.50}                                                               & \multicolumn{1}{c|}{39.76}                                                                           & \multicolumn{1}{c|}{3.52}                                                                             & 0.49                                                                        \\ \hline
\multicolumn{1}{c|}{SGPT-1.3B}            & \multicolumn{1}{c|}{1.3B}       & \multicolumn{1}{c|}{13213.93}                                                              & \multicolumn{1}{c|}{39.39}                                                                           & \multicolumn{1}{c|}{3.19}                                                                             & 1.44                                                                        \\ \hline
\multicolumn{6}{c}{Biomedical Domain}                                                                                                                                                                                                                                                                                                                                                                                                                                 \\ \hline
\multicolumn{1}{c|}{BiomedBERT}           & \multicolumn{1}{c|}{0.1B}       & \multicolumn{1}{c|}{732.14}                                                                & \multicolumn{1}{c|}{3.20}                                                                            & \multicolumn{1}{c|}{0.42}                                                                             & 0.34                                                                        \\ \hline
\multicolumn{1}{c|}{MedCPT-Query}         & \multicolumn{1}{c|}{0.1B}       & \multicolumn{1}{c|}{732.14}                                                                & \multicolumn{1}{c|}{3.16}                                                                            & \multicolumn{1}{c|}{0.45}                                                                             & 0.24                                                                        \\ \hline
\multicolumn{1}{c|}{BMRetriever-410M}     & \multicolumn{1}{c|}{0.4B}       & \multicolumn{1}{c|}{9929.08}                                                               & \multicolumn{1}{c|}{12.09}                                                                           & \multicolumn{1}{c|}{1.18}                                                                             & 0.47                                                                        \\ \hline
\multicolumn{1}{c|}{BMRetriever-1B}       & \multicolumn{1}{c|}{1B}         & \multicolumn{1}{c|}{15846.62}                                                              & \multicolumn{1}{c|}{28.42}                                                                           & \multicolumn{1}{c|}{2.19}                                                                             & 1.40                                                                        \\ \hline
\multicolumn{1}{c|}{BiCA-small}           & \multicolumn{1}{c|}{0.03B}      & \multicolumn{1}{c|}{296.63}                                                                & \multicolumn{1}{c|}{1.75}                                                                            & \multicolumn{1}{c|}{0.47}                                                                             & 0.14                                                                        \\ \hline
\multicolumn{1}{c|}{BiCA-base}            & \multicolumn{1}{c|}{0.1B}       & \multicolumn{1}{c|}{732.90}                                                                & \multicolumn{1}{c|}{3.35}                                                                            & \multicolumn{1}{c|}{0.45}                                                                             & 0.29                                                                        \\ \hline
\multicolumn{1}{c|}{BioHiCL-base (ours)} & \multicolumn{1}{c|}{0.1B}       & \multicolumn{1}{c|}{733.32}                                                                & \multicolumn{1}{c|}{3.50}                                                                            & \multicolumn{1}{c|}{0.63}                                                                             & 0.24                                                                        \\ \hline
\multicolumn{1}{c|}{BioHiCL-large (ours)}  & \multicolumn{1}{c|}{0.3B}       & \multicolumn{1}{c|}{1691.68}                                                               & \multicolumn{1}{c|}{9.99}                                                                            & \multicolumn{1}{c|}{1.20}                                                                             & 0.56                                                                        \\ \hline
\end{tabular}%
}
\caption{Latency and GPU memory usage for experiments on the SciFact information retrieval dataset, following the BEIR benchmark format. Batch size during encoding process is set as 16. Metrics include average corpus and query encoding latency, average retrieval latency, and peak GPU memory usage. Avg Corpus Encoding Latency (ms/doc): Time to encode one document from the corpus. Peak GPU Memory (MB) Usage: Peak GPU RAM during query and corpus encoding. Avg Query Encoding Latency (ms/query): Time to encode one query. Avg Retrieval Latency (ms/query): Time to retrieve top-K documents per query.}

\label{latent_analysis}
\end{table*}

Table~\ref{latent_analysis} reports the latency and GPU memory consumption of the evaluated models. We measure three metrics: average corpus encoding latency (ms per document), average query encoding latency (ms per query), and average retrieval latency (ms per query). Peak GPU memory usage (MB) is also recorded during query and corpus encoding.

\paragraph{Latency and Computational Efficiency of BioHiCL.} 
Our BioHiCL-base model achieves low latency with only 3.50 ms per document for corpus encoding and 0.63 ms per query for query encoding, which is comparable to other small models such as BiCA-small (3.35 ms/doc, 0.45 ms/query). BioHiCL-base also achieves low retrieval latency in its parameter class (0.24 ms/query), demonstrating efficient performance in deployment.

Similarly, BioHiCL-large maintains promising latency-performance among medium-sized models (0.3B parameters), with corpus encoding at 9.99 ms/doc and query encoding at 1.20 ms/query, which is competitive with models of similar size. Its retrieval latency of 0.56 ms/query remains moderate, ensuring efficient search.

Both BioHiCL-base and BioHiCL-large are memory efficient (~733 MB for the base model and ~1.7 GB for the large model), consistent with other models of similar size. This ensures practical deployment of the models without excessive GPU requirements.

\paragraph{Comparison Across Model Sizes.} 
Smaller and medium models (0.1B and 0.3B) exhibit low corpus and query encoding latency, while large models (>=1B) experience substantially higher latency, particularly in corpus encoding (up to ~40 ms/doc for gtr-t5-xl and SGPT-1.3B). 

Peak GPU memory usage scales approximately with model size. Small models (<=0.1B) require ~730 MB, medium models (0.3B) require ~1.7 GB, while large models (>0.3B) demand 8 to 13 GB. 

Large models incur higher latency and require more GPU resources, with corpus encoding exceeding 39 ms/doc and peak GPU memory usage ranging from 8 GB to over 13 GB. While large models may achieve strong retrieval performance, their computational demands make them less practical for low-latency or resource-constrained settings.

\end{document}